\documentclass[12pt]{article}
\usepackage{tikz}
\usepackage{longtable}
\usepackage{graphicx} 
\usepackage{fullpage}
\usepackage{float}
\usepackage{xcolor}
\usepackage{comment}
\usepackage[margin=1in]{geometry}
\usepackage{longtable}
\usepackage{array}
\usepackage{booktabs}
\usetikzlibrary{shapes.geometric, arrows, positioning, calc}
\usepackage{float}

\author{Sarah Hladikova, Yuling Wang, Andreia Martinho }

\date{}
\usepackage{csquotes}
\newcommand{\Comments}{1}
\newcommand{\mynote}[2]{\ifnum\Comments=1\textcolor{#1}{#2}\fi}

\begin{document}

\title{The Third Moment of AI Ethics: Developing Relatable and Contextualized Tools}



\maketitle

\begin{abstract}
  Artificial intelligence (AI) ethics has gained significant momentum, evidenced by the growing body of published literature, policy guidelines, and public discourse. However, the practical implementation and adoption of AI ethics principles among practitioners has not kept pace with this theoretical development. Common barriers to adoption include overly abstract language, poor accessibility, and insufficient practical guidance for implementation. Through participatory design with industry practitioners, we developed an open-source tool that bridges this gap. Our tool is firmly grounded in normative ethical frameworks while offering concrete, actionable guidance in an intuitive format that aligns with established software development workflows. We validated this approach through a proof of concept study in the United States autonomous driving industry.
\end{abstract}

\section{Introduction}

The AI Ethics community responded to the call for normative guidance of AI with a multitude of soft governance mechanisms, such as principles, frameworks, and guidelines \cite{morley2021operationalising,hagendorff2022virtue,munn2023uselessness}. These documents were critical for the development of a shared normative foundation about AI, revolving around core principles modeled after the classic Medical Ethics Principles. Meta-analyses of AI ethics guidelines show that there is a consensus that AI technologies should be beneficial to people and the environment (\textit{beneficence}); robust and secure (\textit{non-maleficence}); respectful of human values (\textit{autonomy}); fair (\textit{justice}); and explainable and accountable (\textit{explainability}) \cite{jobin2019global,floridi2022unified,schiff2020s,morley2020initial,martinho2022empirical,mittelstadt2019principles}. 

The early efforts in developing AI ethics guidelines had limited effect on software development practice \cite{vakkuri2020current}. Hence, the AI Ethics community focused on operationalizing these principles and guidelines \cite{prem2023ethical,ashok2022ethical, morley2021operationalising, morley2020initial, john2022reality, ayling2022putting, vakkuri2021eccola, stix2021actionable, wilson2022sustainable, georgieva2022ai, johnson2021towards, eitel2021beyond, theodorou2020towards}. The Morley Typology, a staple in this so-called \textit{from what to how} moment of AI Ethics, is a comprehensive framework that matches available methods and tools to the core AI Ethics principles (\textit{Beneficence, Non-Maleficence, Justice, Autonomy, and Explainability}), and to stages in the algorithm development pipeline \textit{(Business Development; Design; Training and Test; Prototyping; Testing; Development; Monitoring}) \cite{morley2020initial}. 

However, it seems that these principles and tools have yet to be integrated into technology R\&D practices \cite{hagendorff2022virtue}. 
Several reports indicate their minimal impact on professional practices \cite{vakkuri2020current,orr2020attributions,johnson2021towards,vakkuri2022software,munn2023uselessness}. One factor contributing to this challenge is that these normative mechanisms are presented as external entities with respect to particular domains of science and technology, thus lacking \textit{relatability}  \cite{morley2021operationalising}.

We consider that, as AI continues to become more embedded in different aspects of society, the AI Ethics community should focus on improving the relatability and context specificity of the normative tools. We are aligned with scholars who have called for AI Ethics to develop tools that go beyond general guidelines for professional and ethical practice, reflect normative diversity, do not require a deep background in philosophy, and ensure all stakeholders are engaged and involved in design decisions \cite{morley2021operationalising}.

In this research, we first reflect on the contemporary challenges of AI Ethics and subsequently describe an AI Ethics tool that we developed in what we propose to be the third moment of AI ethics. The tool is grounded in the current normative literature of AI Ethics, yet it is also practical. We included a proof of concept study in the Autonomous Driving industry and requested input from practitioners working in autonomous driving startups in the United States.

\section{ The Challenges of AI Ethics}

\subsection{AI Ethics as a Branch of Applied Ethics}

Ethics explores matters of right and wrong, reflecting the moral spectrum of a particular concept, policy, or technology \cite{martinho2022empirical}. The normative debates welcome disagreement, speculation, and abstraction, thus allowing both diversity of thought and serious ethical consideration \cite{mittelstadt2019principles,martinho2022empirical}. These debates slowly build a robust, rich, and diverse normative foundation, which eventually serves as the basis for governance and regulatory mechanisms. 

AI Ethics explores the ethical implications of AI technologies. It scrutinizes the moral issues arising from the design, development, deployment, and use of AI systems. Like other branches of Applied Ethics, such as Medical Ethics, the normative principles and tools may be used to reflect on complex moral dilemmas, but they also aim to guide the practice (i.e. technology development and innovation).

The community working in this space has focused mainly on translating and incorporating normative elements through the design pipeline. These issues are deeply embedded in the larger societal context, namely in domains such as Transportation \cite{martinho2021ethical}, Healthcare \cite{martinho2021healthy}, or Justice \cite{martinho2024surveying}. The socio-technical nature of AI prompts ethicists to examine the role of software developers and other practitioners in the outcomes of the technology.

\subsection{The Challenges of AI Ethics in the current AI Paradigm}

AI Ethics faces unique challenges related to the \textit{urgency of normative guidance}, \textit{multi-purpose nature of AI}, and
\textit{multitude of stakeholders} \cite{martinho2022empirical}.
The recent state of affairs in AI is characterized by the fast development and deployment of these technologies, as well as an urgent need for practical and operational normative guidance, resulting in an \textit{AI Ethics Boom}, \cite{CORREA2023100857} where many academic, government, and industry organizations started publishing normative guidance to AI. These efforts were predominantly concerned with the principles that ought to guide the \textit{modus operandi} of practitioners, however, as mentioned earlier, there was little to none guidance on the implementation of such principles into practice \cite{CORREA2023100857}. 

Another challenge for AI Ethics is related to the fact that modern AI is a multi-purpose
technology. AI is currently applied to several societal domains, such as Transportation,
Healthcare, or Justice \cite{martinho2022empirical}. AI Ethics needs to consider how risks, conflicting rights and interests, and social preferences vary in different contexts \cite{morley2021ethics}. Just as in other Applied Ethics domains, namely Medical Ethics or Business Ethics, there are many distinct issues that require rigorous examination and scrutiny.

As a result of the multi-purpose nature of AI, there are many stakeholders involved in the development of various AI-powered technologies. Organizations such as Google, Amazon, Facebook, or OpenAI, have shown a strong interest in AI Ethics and have developed several guidelines \cite{hagendorff2020ethics,Ali_2023}. Traditionally, the Ethics community is hesitant about industry-led initiatives regarding Ethics and often dismisses these initiatives as attempts to shape the normative conversations to their own interests \cite{morley2020initial, jobin2019global}. 
However, it may also be argued
that these organizations produce guidelines as a response to the Ethics communities (as well as, perhaps, concerns and criticism in popular culture and media). The Ethics work is rarely \textit{ready to use} as it is often abstract, speculative, and intricate. 

A challenge for the AI Ethics researchers is to operationalize their work and communicate effectively with other communities in the AI space  \cite{martinho2022empirical}. Past experience with other applied fields of Ethics (e.g., Medical Ethics) shows that it is possible to operationalize Ethics from abstract ethical principles successfully, thus lending credence to the AI Ethics effort \cite{morley2021ethics}. 
This research builds on the premise that AI Ethics needs to focus on improving the relatability and contextualization of its normative tools\cite{martinho2022empirical}.

\subsection{The Three Moments of AI Ethics}

The AI Ethics work developed in recent years meets the traditional expectations in terms of richness, controversy, and diversity of thought \cite{martinho2022empirical}. There have been numerous contributions, including guidelines, white papers, or company statements of commitment to ethical design (\cite{eu2019ethics, berkman2019ai, eu2020whitepaper}. Still, the research on the preparedness of practitioners to face ethical dilemmas in the design and development of artificial intelligence implies a disconnect between established ethical norms and practice \cite{brightman2018applying}. To better understand the disconnect between research and practice, we revisit the three moments in the field of AI ethics.

In the first moment of AI Ethics, as mentioned earlier,  a multitude of soft governance mechanisms were developed by the scientific community, governments, private companies, and non-governmental organizations \cite{jobin2019global, hagendorff2020}. The subject of ethical inquiry was the notion of alignment between moral values and the machine objectives \cite{gabriel2020}, and much of the research efforts focused on identifying the principles and values that ought to guide the software engineering practice \cite{alma991634288204371, floridi2018ai4people}. However, ethics principles have been deemed as abstract, not relatable, and yet to be integrated into technology practices, sparking concerns about the success of the AI Ethics endeavor. 

Moreover, there is an ongoing mistrust that some of the organizations that contributed to this normative corpus, namely private organizations involved in the development of AI technologies, will go beyond \textit{ethics washing} and 
implement ethical practices voluntarily. When Ethics is integrated into organizations, there are concerns that it is used merely as a marketing strategy with little impact when it
comes to decisions \cite{hagendorff2020ethics, orr2020attributions}.  A study that surveyed AI practitioners about their perceived impact of AI Ethics guidelines reported that the effectiveness of such guidelines or ethical codes is extremely low and that they do not change the behavior of professionals from the tech community \cite{johnson2021towards}.

In a second moment, the AI Ethics community focused on translating abstract principles to more specific practice-oriented guidelines (the so-called \textit{from what to how} phase in AI Ethics \cite{morley2020initial}. However, there is little evidence of the impact of these tools and methods on 
the governance of AI \cite{morley2020initial}. 
The research shows that ethical guidelines do not translate into practice. Studies have reported that practitioners do not see ethics as a concern or challenge during development of Machine Learning models, and do not account for ethics in their practices, even though they might find ethical guidelines overall useful for their organizations or care about ethics on a personal level \cite{vakkuri2020current,johnson2021towards,munn2023uselessness}. 

The research shows that there is a gap between the research of ethics and technology, and its application in the engineering field \cite{karim2017ethical}. Engineers perceive ethics as \textit{system of barriers and constraints} and the complex language of ethics prevents its application and deployment throughout the innovation process \cite{byers2021empowering}.

As mentioned earlier, Morley et al. identified the tools and methods already available to guide AI practitioners on core issues of AI Ethics and plotted these methods and tools in a typology, matching them to ethical principles and stages in the algorithm development pipeline. They reported that numerous tools and methodologies exist to assist practitioners in realizing Ethical AI, but the vast majority are severely limited in terms of usability \cite{morley2020initial}. 

In the third moment of AI Ethics, the community needs to focus on developing tools that are relatable and contextualized. We align with Morley at al., AI Ethics requires an approach that (i) goes beyond general guidelines for professional and ethical practice, (ii) embodies a tool set that does not require a deep background in philosophy, (iii) reflects the normative status of ethical reasoning, (iv) ensures all stakeholders are engaged and involved in design decisions rather than simply consulted about them; and (v) reflects the language and tensions in particular fields \cite{morley2021operationalising}.

In this research, we make a contribution to the third moment of AI Ethics. We built on the Morley Typology and developed a normative tool that is theoretically grounded, diverse, and practical, and can be contextualized in different domains. By translating abstract moral values such as fairness or transparency into specific applicable items that can help guide the process of technological development, we provide a tool and language that resonates with the practitioners and empowers them to engage in complex conversations over the socio-technical nature of their work. 

\section{The AI Ethics Tool}

\subsection{Review of the Morley Typology}
The AI Ethics tool is inspired by the Morley Typology \cite{morley2020initial}.  This theoretical framework matches N= 106 available methods and tools reported in the literature to the core AI Ethics principles (\textit{Beneficence, Non-Maleficence, Justice, Autonomy, and Explainability}), and to phases in the algorithm development pipeline (\textit{Business Development; Design; Training and Test; Prototyping; Testing; Development; Monitoring}). We first reviewed the literature associated with each phase and excluded the methods and tools no longer available. We used this compressed Morley Typology as our starting point for the development of the AI Ethics tool, but we also reviewed and included methods and tools beyond the Typology that were also aligned with such mandates. 
The review of Morley Typology served as our foundational starting point, focusing on its categorization of AI ethics tools across development phases and value representations. We systematically analyzed each category and tool mentioned in the typology, verifying their current availability, maintenance status, and updates since the original publication. This verification process was crucial for understanding the evolving landscape of AI ethics tools and identifying gaps in existing resources.
Our analysis expanded beyond the original typology through comprehensive literature review using Google Scholar and standard search engines, which led to the discovery of additional frameworks such as the NIST AI Risk Management Framework \cite{tabassi2023ai}. We classified these new resources according to Morley's original categories while adding new dimensions for normative ethics foundations and key elements, creating a more comprehensive inventory that reflected current developments in AI ethics tools. 

\tikzstyle{process} = [rectangle, minimum width=3cm, minimum height=1cm, text centered, draw=black]
\tikzstyle{arrow} = [thick, ->, >=stealth]

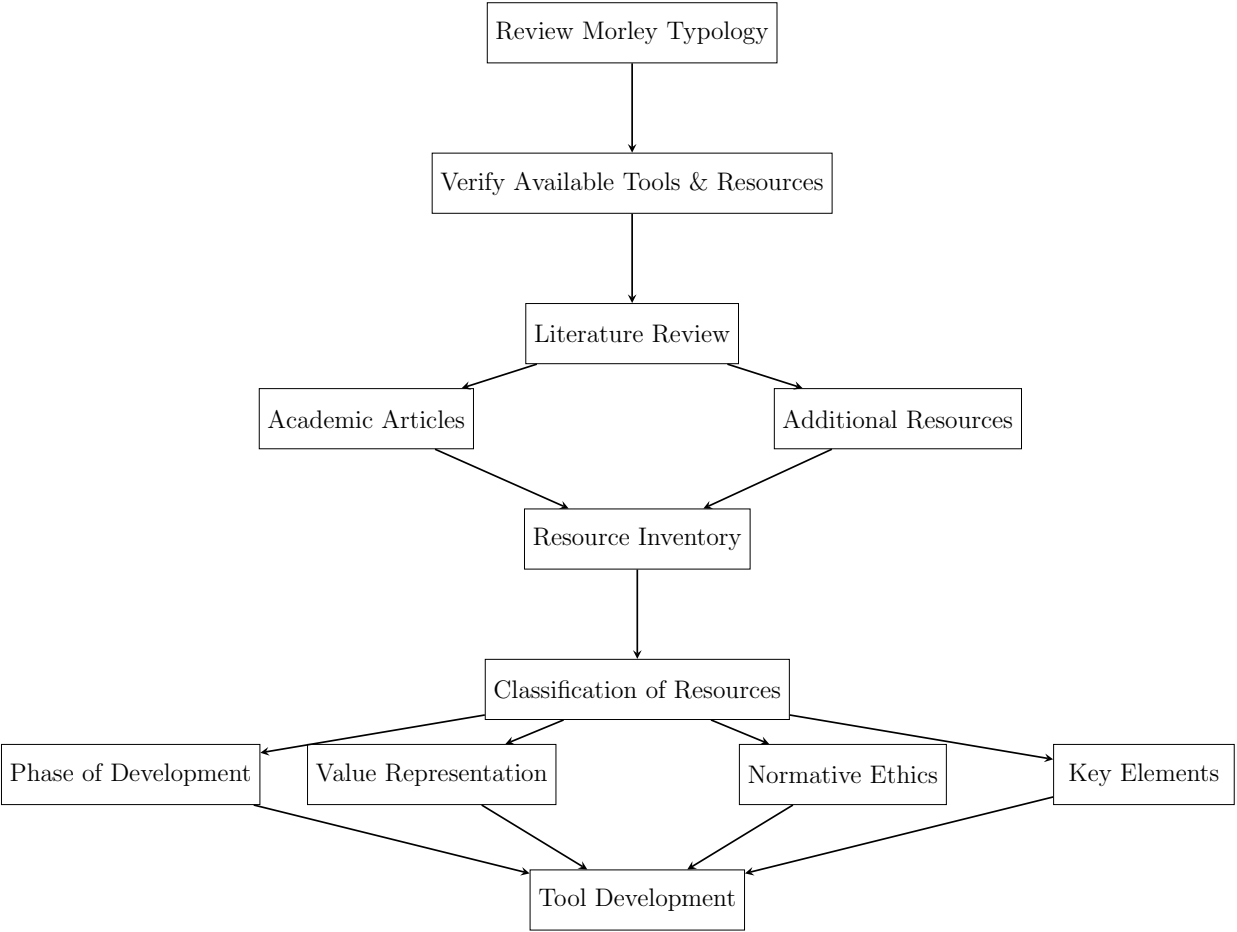
\begin{figure}[htbp]
\centering
\resizebox{\textwidth}{!}{%
\begin{tikzpicture}[node distance=2cm, auto]
    \node (A) [process] {Review Morley Typology};
    \node (B) [process, below of=A, yshift=-0.5cm] {Verify Available Tools \& Resources};
    \node (C) [process, below of=B, yshift=-0.5cm] {Literature Review};
    \node (D) [process, below left of=C, xshift=-3cm] {Academic Articles};
    \node (E) [process, below right of=C, xshift=3cm] {Additional Resources};
    \node (F) [process, below of=D, xshift=4.5cm] {Resource Inventory};
    \node (G) [process, below of=F, yshift=-0.5cm] {Classification of Resources};
    \node (H) [process, below left of=G, xshift=-7cm] {Phase of Development};
    \node (I) [process, below left of=G, xshift=-2cm] {Value Representation};
    \node (J) [process, below right of=G, xshift=2cm] {Normative Ethics};
    \node (K) [process, below right of=G, xshift=7cm] {Key Elements};
    \node (L) [process, below of=G, yshift=-1.5cm] {Tool Development};

    \draw [arrow] (A) -- (B);
    \draw [arrow] (B) -- (C);
    \draw [arrow] (C) -- (D);
    \draw [arrow] (C) -- (E);
    \draw [arrow] (D) -- (F);
    \draw [arrow] (E) -- (F);
    \draw [arrow] (F) -- (G);
    \draw [arrow] (G) -- (H);
    \draw [arrow] (G) -- (I);
    \draw [arrow] (G) -- (J);
    \draw [arrow] (G) -- (K);
    \draw [arrow] (H) -- (L);
    \draw [arrow] (I) -- (L);
    \draw [arrow] (J) -- (L);
    \draw [arrow] (K) -- (L);
\end{tikzpicture}
}
\caption{From Morley Typology Review to the Tool Development}
\label{fig:research-flow}
\end{figure}

\subsection{Software Development}

The development infrastructure centered on GitBook (https://ai-ethics-tool.gitbook.io/ai-ethics-tool), a documentation platform popular for capturing and documenting technical knowledge, such as product documentation, internal knowledge bases, or Application Programming Interfaces. GitBook is integrated with Git version control. This setup enabled seamless two-way synchronization between GitBook and Git branches, facilitating documentation updates while making the tool open source. Changes made in either platform automatically reflected in the other, streamlining the development process.

The technical implementation utilized GitBook's robust editor capabilities to organize content into eight markdown files, each corresponding to different AI validation stages: building, deployment, design, development, ethics, monitoring, testing, and training. The editor supported various formats including text, tables, and code blocks, enhancing content readability. The development process began with initial documentation drafting, followed by content refinement based on research findings. The team then focused on web design planning to ensure effective delivery of practical normative guidance, culminating in the implementation of layout and interface elements optimized for accessibility and user engagement.

\begin{figure}[htbp]
   \centering
   \includegraphics[width=0.8\textwidth]{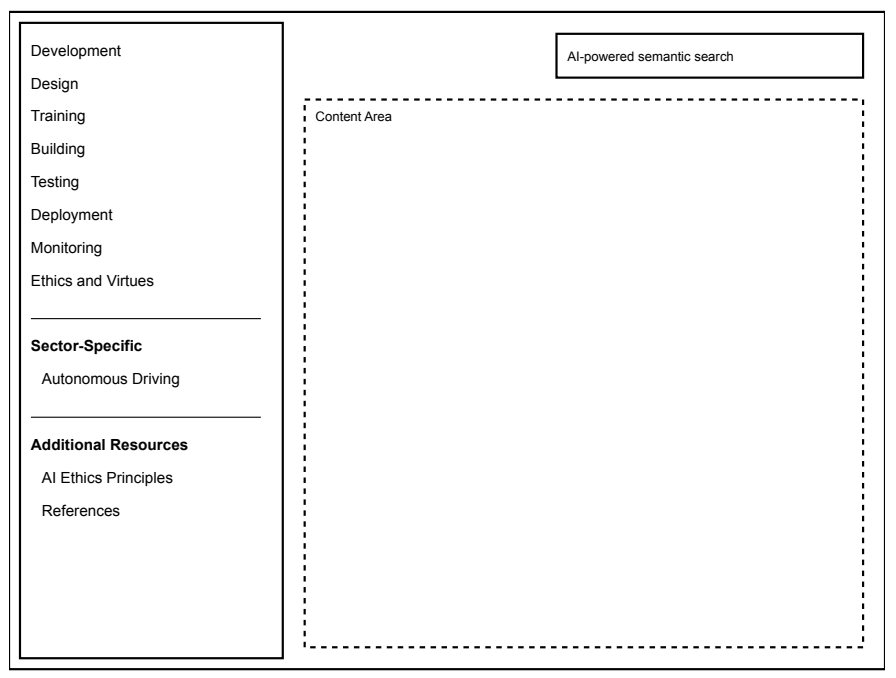}
   \caption{The AI Ethics Tool Schematic Representation}
   \label{fig:unique-label}
\end{figure}

The tool is divided into sections reflecting the algorithm development pipeline, adjusted from the Morley typology \cite{morley2020initial} as (i) Development (ii) Design (iii) Training (iv) Building (v) Testing (vi) Deployment (vii) Monitoring and (viii) Fostering Ethics and Virtues. Our technology-agnostic, open source model under the MIT license allows practitioners to create their own version suitable for their specific needs. As an example of further operationalization, we developed a tab specifically aimed at autonomous vehicles industry. To further improve the intuitive interface, we implemented an AI-powered semantic search, which allows users to search in natural language and returns answers summarizing the content and linking directly to relevant information within the website.

\begin{figure}[htpb]
   \centering
   \includegraphics[width=0.8\textwidth]{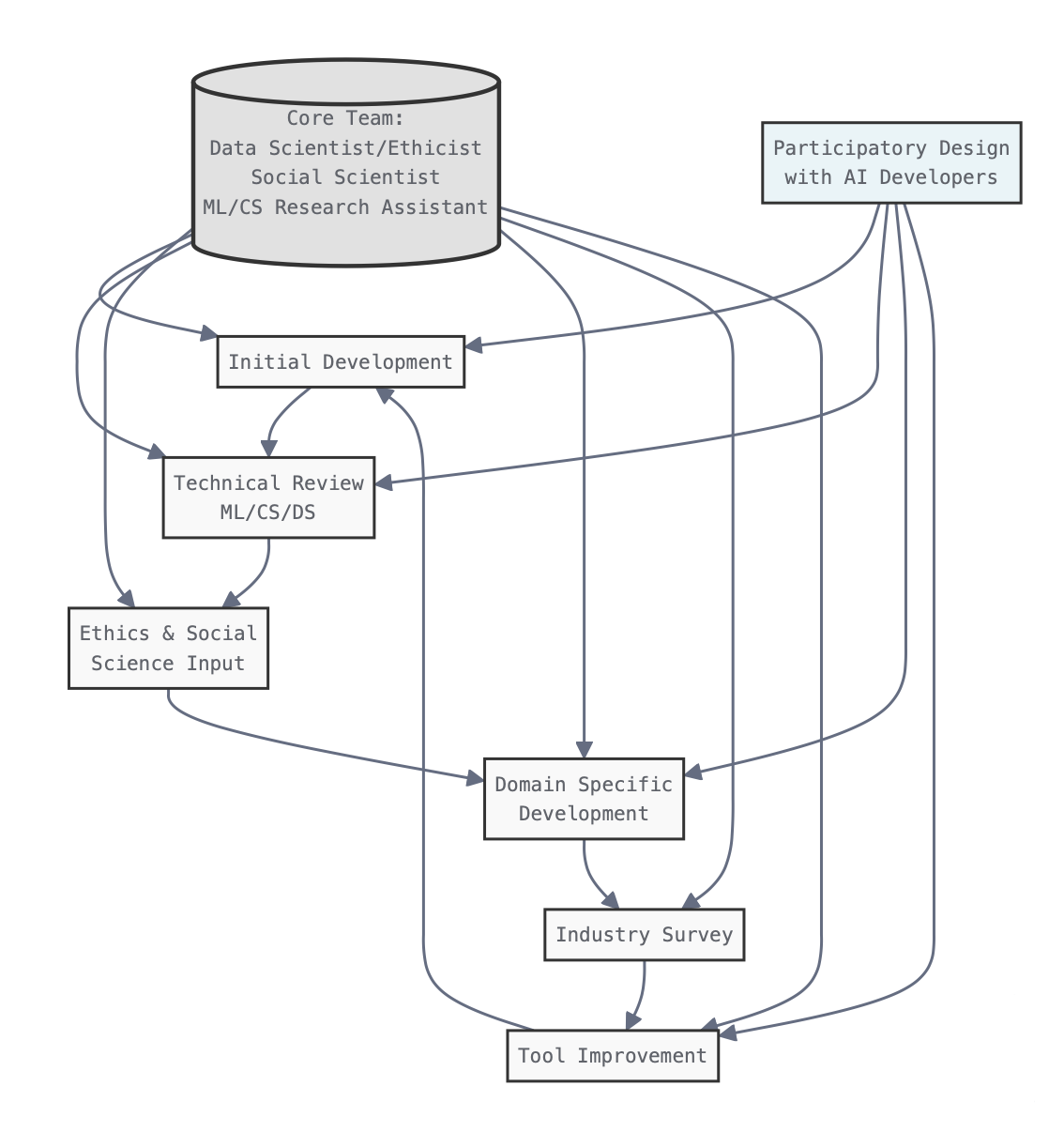}
   \caption{AI Ethics Tool Development Cycle: Proof of Concept Study with Continuous Stakeholder Engagement}
   \label{fig:unique-label}
\end{figure}

\subsection{Testing the Tool}
\subsubsection{Proof of Concept Study on Autonomous Driving}
For the purpose of testing the tool and assessing its relatability, we developed an empirical study focusing on the Autonomous Driving industry.  We consider that this industry makes a compelling audience for our study, as the Autonomous Vehicle (AV) has been widely used in the AI Ethics literature to illustrate the ethical trade-offs in traffic situations that entail the distribution of risk \cite{bonnefon2016social,bonnefon2019trolley, martinho2021ethical}. The particular aims of the proof of concept study are  (i) to engage practitioners in the AV community; (ii) to assess the relatability of the tool; (iii) to receive feedback and make improvements in the tool; and (iv) to follow up on previous studies and investigate whether this approach has the potential to be useful for practitioners.

\subsubsection{Survey}
We developed a survey using the Qualtrics (https://www.qualtrics.com.), a software for creating and implementing online surveys. The study was submitted to the Tufts University Institutional Review Board (IRB) and received an exempt determination. The final version of the survey features thirteen groups of questions (Supplementary Materials). The first group of questions relates to (i) Professional Role \& Experience; (ii) Knowledge of Ethics; (iii) Perspectives about AI Ethics; (iv) AI Ethics in Industry Projects; (iv) Relatability, Usefulness, and Feedback on the Tool.

\subsubsection{Results}

We deployed the survey via email to (N=40) Autonomous Driving companies in January 2024 (Supplementary Materials). The engagement in the survey was low and we recorded only nine responses (seven with no missing data). The data was stored in Qualtrics. The results shown below relate to the characterization of the study participants, their (self-perceived) knowledge and perspectives about (AI) Ethics in general \cite{penuel2017how} and in their projects, and also reliability and usefulness of the tool. The comments and feedback provided by the participants are included in the Supplementary Materials.

\section{Discussion \& Conclusions}

As a novel Applied Ethics field, AI Ethics has experienced rapid development, reflecting the emerging need for normative discussion around this transformative technology. Despite the richness of AI Ethics research, there are concerns about its success in realizing normative endeavors, especially in light of the latest developments in Generative AI \cite{hagendorff2024mapping}. Empirical research shows that even when ethical guidelines resonate with developers, these normative efforts do not translate into practices \cite{vakkuri2020current,johnson2021towards,munn2023uselessness,johnson2021towards,byers2021empowering}. 
 We contend that AI Ethics needs to move to its Third Moment, in which it is contextualized within particular scientific and technology domains, in order to reflect the language, tensions, and priorities of such domains and, therefore, improve relatability. Understanding and utilizing practices and tools that practitioners use further enhances their integration into the operational processes of a specific team or organization. 

We developed a theoretically grounded yet practical AI Ethics tool that combines numerous translational tools and methods to assist practitioners in incorporating Ethics into their practices. We included a domain-specific tab related to Autonomous Driving. We also developed a small empirical study to test this tool by surveying practitioners currently working in the Autonomous Driving space. Although this study has clear limitations, related to the small sample of practitioners who participated in the study (N=9), we received a good amount of feedback that allow us to further improve the tool.  

Our results are aligned with previous empirical studies indicating that practitioners acknowledge the importance of AI Ethics, but do not account for it in their practices \cite{vakkuri2020current,johnson2021towards,munn2023uselessness}. The participants in our empirical study indicate they have some knowledge about Ethics, AI Ethics, and Autonomous Driving Ethics but are somewhat neutral about their perspectives on AI Ethics and AI Ethics researchers. They also signal that Ethics comes up more often in the later phases of the innovation pipeline. Regarding our tool, the participants considered that it is relatable and the flow of the tool is adequate. The participants also provided extensive written feedback regarding potential barriers for adoption, that will be incorporated into the next iteration of the tool. 

Through the analysis of the comments, it is clear that the practitioners who provided feedback on the tool still wish for more detail, precise instructions, and examples. Participants often mention that the language is still high level, thus reinforcing once again the need for relatable and contextualized AI Ethics tools.

As AI Ethics moves forward, the community needs to reflect on the best ways to improve the relatability and context-specificity of the normative tools, namely through collaborative endeavors, so that different communities can further engage and contextualize AI Ethics. The private sector has a dominant position in much of the research and development of AI. Industry self-regulation can help to ensure that AI systems are developed and used in a way that aligns with the values and interests of society. However, self-regulation may be misinformed or lack crucial perspectives to address the potential risks and impacts of AI, especially in cases where the technology is used in areas such as healthcare or public safety. Research-grounded interventions such as the introduced AI Ethics Tool provide more accessible venue for practitioners to meaningfully engage in conversations around ethics and the normative aspects of their work, while utilizing existing processes and reflecting familiar practices.

\bibliographystyle{abbrv}
\bibliography{data_ethics}

\appendix
\section{Tools and resources from the Morley Typology for the AI Ethics Tool}

\setlength\LTleft{0pt}
\setlength\LTright{0pt}

\begin{longtable}{>{\raggedright\arraybackslash}p{0.12\textwidth}>{\raggedright\arraybackslash}p{0.13\textwidth}>{\raggedright\arraybackslash}p{0.20\textwidth}>{\raggedright\arraybackslash}p{0.35\textwidth}>{\raggedright\arraybackslash}p{0.20\textwidth}}

\toprule
\textbf{Phase} & \textbf{Value} & \textbf{Key elements} & \textbf{Algorithm} & \textbf{Ethics} \\
\midrule
\endhead

\bottomrule
\endfoot

Development & Beneficence & Reflecting on consequences & Consequence scanning, Ethics cards, Wellcome Data Labs & Consequentialism \\

Development & Beneficence & Foster engagement in AI Ethics & Involve & Virtue Ethics \\

Development & Non-Maleficence & Critical conditions to avoid harm & An ethical framework for evaluating experimental technology (16 conditions) & Deontology \\

Development & Justice & Compatibility with universally accepted principles & Algorithmic Accountability and Public Reason & Contractualism \\

Development & Justice & Determining social impact of algorithm (this a self-contained framework) & Principles for Accountable Algorithms and a Social Impact Statement for Algorithms & Deontology/ Principles \\

Development & Explainability & Need for documentation and metadata & FAIR & Deontology \\

Design & Beneficence & Full description of the project/ algoritms/ responsibility for government AI projects & Data Ethics Decision Aid (DEDA): a dialogical framework for ethical inquiry of AI and data projects in the Netherlands & Virtue/Value Sensitive \\

Design & Explainability & Research must be reproducible & FAIR, The Turing way & Deontology/ Principles \\

Training/ Testing & Non-Maleficence & Data should be protected through privacy, minimization and creation of synthetic data & Data minimizers, Hazy, ONS, Tensor Flow Privacy & Value Sensitive Design \\

Training/ Testing & Justice & Datasets should be well described, anonymized, and GDPR compliant & Dataset nutrition label, ICOa, ICOb, Care Principles for Indigenous Data & Deontology/ Principles \\

Training/ Testing & Justice & Gender stereotypes should be removed & Tolga & Value Sensitive Design \\

Training/ Testing & Justice & Datasets should be scrutinized for Fairness & IBM Fairness & Value Sensitive Design \\

Building & Non-Maleficence & Ensure Privacy & Open Mined & Value Sensitive Design \\

Building & Justice & Ensure Privacy \& Fairness & Open Mined, Fairlearn, Confidence-based approach, IBM 360 & Value Sensitive Design \\

Building & Explainability & Ensure Explainability and Interpretability & Case-based reasoning, Interpretability resources, Counterfactual Fairness, XAI & Value Sensitive Design \\

Testing & Beneficence & Investigate the relationship between prediction and decision & Human Decisions and Machine Predictions & Value Sensitive Design \\

Testing & Non-Maleficence & Ensure reproducibility & Turing way & Value Sensitive Design \\

Testing & Justice & Check bias and fairness & Certifying and removing disparate impact, IBM 360, Fairness Experimental Tech & Value Sensitive Design \\

Testing & Explainability & Model should be interpretable & Quantitative Input Influence, SHAP, Microsoft -- InterpretML, LIME, DeepLift XAI, Visual interpretations & Value Sensitive Design \\

Deployment & Non-Maleficence & Foster transparency, accountability, and procurement & AI Now Institute Algorithmic Accountability Policy Toolkit, AI-RFX Procurement Framework, OPAL, Machine Learning Reproducibility Checklist & Value Sensitive Design/ Virtue Ethics \\

Deployment & Autonomy & N/A & & Value Sensitive Design \\

Deployment & Justice & Safeguards must be in place for automated scoring systems & The Scored Society: Due process for automated predictions & Value Sensitive Design \\

Deployment & Explainability & Foster transparency & AI Now Institute Algorithmic Accountability Policy Toolkit & Value Sensitive Design \\

Deployment & Explainability & Trusting suppliers & Factsheets & Value Sensitive Design \\

Deployment & Explainability & Document metadata & Datasheets for Datasets, Model Cards for Model Reporting & Value Sensitive Design \\

Deployment & Explainability & Uncover algorithmic impact & Algorithmic Impact Assessments: A practical Framework for Public Agency Accountability & Value Sensitive Design \\

Monitoring & Non-Maleficence & Create defense mechanisms & Adversarial robustness toolkit & Value Sensitive Design \\

Monitoring & Explainability & Protect not only data but also inferences & A right to reasonable inferences: re-thinking data protection law in the age of big data and AI & Value Sensitive Design \\

\end{longtable}

\section{Survey Results}
	\begingroup 
		\def\arraystretch{1.5} 
		\begin{table} [H]
			\centering
			\footnotesize
			
			\begin{tabular}{lcc} 
				\hline \hline
				\textbf{Professional Role}& \textbf{ Participants (N)}  \\
    Software Development & 5 \\

    Product Management & 1\\
        R\&D & 3\\ \\

				\textbf{Experience}& \textbf{ }  \\
Entry Level (0-3 years) & 2 \\
Intermediate (3-7 years) & 3 \\
Mid-Level (8-15 years) &  3 \\
Senior-level (15+ years) & 1 \\
			
				\hline \hline
				
			\end{tabular}
            \caption {\label{tab:participants} Professional Role and Experience  } 
		\end{table}
        
		\endgroup

	\begingroup
		\def\arraystretch{1.5}
		\begin{table}[htbp]
			\centering
			\footnotesize

			\begin{tabular}{lc}
				\hline \hline
				\textbf{Ethics} & 62\\

				\textbf{AI Ethics}  & 60\\

    		\textbf{AV Ethics} & 63\\

				\hline \hline
				
			\end{tabular}
            \caption {\label{tab:participants}(Self-Perceived) Mean Knowledge of Ethics [0-100]  } 
		\end{table}
		\endgroup

	\begingroup 
		\def\arraystretch{1.5} 
		\begin{table} [htbp]
			\centering
			\footnotesize

			\begin{tabular}{lc}
				\hline \hline
\textbf{Statement} & \textbf{Participants}\\
\hline
\textbf{AI Ethics addresses questions that help my team make better decisions} \\
Somewhat agree & 6 \\
Neither agree nor disagree & 3 \\

\textbf{AI Ethics researchers live in an ivory tower isolated from practice}\\
Somewhat agree & 4 \\
Neither agree nor disagree & 5 \\

\textbf{ The AI Ethics claims are trustworthy}\\
Somewhat agree & 4 \\
Neither agree nor disagree & 5 \\

				\hline \hline
				
			\end{tabular}
            \caption {\label{tab:participants} Perspectives About AI Ethics  }
		\end{table}
		\endgroup

	\begingroup 
		\def\arraystretch{1.5} 
		\begin{table} [H]
			\centering
			\footnotesize

			\begin{tabular}{lc}
				\hline \hline
\textbf{Phase} & \textbf{Participants}\\
\hline
\textbf{Development} & \\

Sometimes & 5\\
About half the time & 3 \\
Most of the time & 1 \\

\textbf{Design} & \\
Sometimes & 3 \\
About half the time & 3 \\
Most of the time & 1 \\
Always & 2 \\

\textbf{Training} & \\
Never & 2\\
Sometimes & 1 \\
About half the time & 3 \\
Most of the time & 2 \\
Always & 1 \\

\textbf{Building} & \\
Never & 1\\
Sometimes & 1 \\
About half the time & 5 \\
Most of the time & 2 \\

\textbf{Testing} & \\
Sometimes & 2 \\
About half the time & 2 \\
Most of the time & 4\\
Always & 1\\

\textbf{Deployment} &\\
Sometimes & 3\\
About half the time & 1 \\
Most of the time & 4\\
Always & 1 \\

\textbf{Monitoring} & \\
Sometimes & 1 \\
About half the time & 2 \\
Most of the time & 4 \\
Always & 2\\

				\hline \hline
				
			\end{tabular}
            \caption {\label{tab:participants} AI Ethics in Industry Projects  }
		\end{table}
		\endgroup

\begingroup 
		\def\arraystretch{1.5} 
		\begin{table} [htbp]
			\centering
			\footnotesize

			\begin{tabular}{lc}
				\hline \hline
\textbf{Relatability} & \textbf{Participants}\\
Relatable & 9 \\

\textbf{Adoption} & \\

Yes & 3 \\
No & 3 \\

\textbf{Flow} & \\

Yes & 7  \\
No & 2 \\

				\hline \hline
				
			\end{tabular}
            \caption {\label{tab:participants}  Relatability, Usefullness, and Feedback on the Tool  } 
		\end{table}
		\endgroup

\section{Survey Results Analysis}

\subsection{Need for Examples}
\begin{itemize}
    \item Practical implementation examples
    \item Case studies of ethical decisions
    \item Organizational scenarios
    \item Team-level implementation examples
\end{itemize}

\subsection{Implementation Guidance}
\begin{itemize}
    \item \textbf{Prioritization of Actions}
    \begin{itemize}
        \item Distinguish critical vs. nice-to-have steps
        \item Address resource/bandwidth constraints
    \end{itemize}
    \item \textbf{Iterative Processes}
    \begin{itemize}
        \item Model fine-tuning procedures
        \item Edge case handling protocols
    \end{itemize}
    \item \textbf{Team Dynamics}
    \begin{itemize}
        \item Small group decision-making frameworks
        \item Cross-team coordination strategies
    \end{itemize}
\end{itemize}

\subsection{Content Structure}
\begin{itemize}
    \item Convert to conversational tone
    \item Include step-by-step instructions
    \item Reorganize into itemized lists
    \item Add contextual details
\end{itemize}

\subsection{Missing Elements}
\begin{itemize}
    \item Community engagement methodologies
    \item Statistical analysis frameworks
    \item Ethical metric definition guidelines
    \item Code practice recommendations
    \item Input validation procedures
\end{itemize}

\subsection{Section Organization}
\begin{itemize}
    \item Merge overlapping sections (e.g., Training and Building)
    \item Expand monitoring scope
    \item Create dedicated discussion groups section
\end{itemize}

\section{Research Ethics and Social Impact}
\subsection{Ethical Considerations Statement}
This research was conducted with guidance from the IRB office at Tufts University. All survey participants provided informed consent and were informed about data collection, usage, and anonymization procedures. No personally identifiable information was collected. Survey participation was voluntary and unpaid. The study focused on practitioners from autonomous vehicle companies, acknowledging potential selection bias in our sample. We maintained transparency about sample size limitations (N=9) and their impact on generalizability. Our open-source tool development prioritized accessibility and ethical implementation guidelines while avoiding prescriptive requirements that could disadvantage smaller organizations or teams with limited resources.
\subsection{Adverse Impact Statement}
While our research aims to improve AI ethics implementation, we acknowledge potential adverse impacts. Our tool could inadvertently promote superficial compliance rather than meaningful ethical engagement, particularly disadvantaging resource-constrained organizations. The autonomous vehicle industry focus may lead to inappropriate generalization across domains. We address these risks through open-source availability, adaptable implementation guidance, and continuous community feedback mechanisms, while remaining mindful that our tool should complement rather than replace deeper ethical discourse and diverse stakeholder engagement.
\end{document}


\textbf{Development} 

\textit{ I would say that the development phase involves several cycles of design, training, building, and testing steps.}

\textit{This reads like a syllabus / lesson plan. It could be more conversational, like a how-to guide or step-by-step instructions.}

\textbf{Design}
\textit{One missing element might be iterative design.  Often a model is fine-tuned / improved with newer data over time.  How to maintain high ethical practices while potentially mining for edge-case situations may be an interesting point to cover.}

\textit{Provide examples}

\textbf{Training}

\textit{In practice, I don't think there is a big distinction between the Training and Building sections described by this tool. I think the tool would be more concise if there was a way to combine these sections Training as often done as a means to Building.}

\textbf{Testing}

\textit{What do you mean with Conduct regular and sustained engagement with potentially impacted communities. ? The suggested actions section is very comprehensive and most of the things are not possible to do because there is no bandwidth to do all those steps and still produce a model.  I suggest to organize this section in different levels such as critical actions, nice to have and not that important.}

\textbf{Deployment}

\textit{There can more details for the training data. }

\textit{I think the section is a bit general in its questions and may be difficult to actualize when it comes to specific deployments. I also think it would be important to talk about statistical analysis and also in general how to define metrics ethically. Often times metrics are designed to look good but are not really holistic. }

\textit{Outside of the general feedback of examples of the tool in practice, nothing specific. I feel like this section is very comprehensive.}

\textbf{Monitoring}

\textit{Here suggested actions can also be added. for instance, good code practices, adapting coding standards, CIs, tests, etc,}

\textit{Need to talk about the input that user provide}

\textit{I think this section can be expanded with examples. }

\textit{Monitoring should extend beyond feedback of engaged communities and include outreach to communities that may be unknowingly affected. It would also be nice to have strategies for determining which communities may be affected by are not speaking up and how to engage with them.}

\textit{A short sentence in each of the four check boxes might help to provide additional context and details.}

\textit{For your audience, making this part an itemized list would be beneficial.}

\textbf{Fostering Virtue Ethics}

\textit{I also think providing organizational examples, like with a make believe company and org structure, would be good to make this more tangible rather than just being generic ideas in peoples heads }

\textit{Extra attention should be brought to the section relating to discussion groups and should maybe even be its own section. I practice, I believe most decisions about AI research and development happen at the team level, groups of 4-10 people that may dictate where individual manpower is dedicated but are may not be representative of an entire organization. It would be helpful to have some tips on navigating team environments.}

\textit{This reads very general / high-level.  Some examples might help to illustrate the ideas more concretely.}

\textit{Any measures for teams?}